\documentclass[conference]{IEEEtran}
\usepackage{cite}
\usepackage{amsmath,amssymb,amsfonts}
\usepackage{algorithmic}
\usepackage{graphicx}
\usepackage{textcomp}
\usepackage{xcolor}

\def\BibTeX{{\rm B\kern-.05em{\sc i\kern-.025em b}\kern-.08em
    T\kern-.1667em\lower.7ex\hbox{E}\kern-.125emX}}

\begin{document}

\title{A Similarity-based Approach to Random Survival Forests}

\author{\IEEEauthorblockN{1\textsuperscript{st} Yingying Xu}
\IEEEauthorblockA{\textit{Department of Statistics} \\
\textit{ and Actuarial Science}\\
\textit{University of Waterloo}\\
Waterloo, Canada\\
}
\and
\IEEEauthorblockN{2\textsuperscript{nd} Joon Lee}
\IEEEauthorblockA{\textit{Cumming School of Medicine} \\
\textit{University of Calgary}\\
Calgary, Canada\\
}
\and
\IEEEauthorblockN{3\textsuperscript{rd} Joel A. Dubin}
\IEEEauthorblockA{\textit{Department of Statistics} \\
\textit{ and Actuarial Science}\\
\textit{University of Waterloo}\\
Waterloo, Canada\\
}
}

\maketitle

\begin{abstract}
Predicting time-to-event outcomes in large databases can be a challenging but important task.  One example of this is in predicting the time to a clinical outcome for patients in intensive care units (ICUs), which helps to support critical medical treatment decisions. In this context, the time to an event of interest could be, for example, survival time or time to recovery from a disease/ailment observed within the ICU. The massive health datasets generated from the uptake of Electronic Health Records (EHRs) are quite heterogeneous as patients can be quite dissimilar in their relationship between the feature vector and the outcome, adding more noise than information to prediction. In this paper, we propose a modified random forest method for survival data that identifies similar cases in an attempt to improve accuracy for predicting time-to-event outcomes; this methodology can be applied in various settings, including with ICU databases. We also introduce an adaptation of our methodology in the case of dependent censoring. Our proposed method is demonstrated in the Medical Information Mart for Intensive Care (MIMIC-III) database, and, in addition, we present properties of our methodology through a comprehensive simulation study. Introducing similarity to the random survival forest method indeed provides improved predictive accuracy compared to random survival forest alone across the various analyses we undertook.
\end{abstract}


\section{Introduction}

% Lee, Maslove \& Dubin (2015) \cite{lee2015personalized}
Electronic Health Records (EHRs) have generated health data sets that provide rich and diverse information for modeling and prediction. Survival analysis has been essential in clinical and epidemiological studies, and both parametric and semiparametric modeling have been utilized in the literature (e.g., \cite{b1}). Especially with big datasets, patients can be heterogeneous, which pose challenges to accurate prediction of outcomes of interest. Conditioning on a more relevant subset where the cases are more similar to the point of prediction might improve prediction accuracy. Similarity-based prediction has been focused upon for other types of responses, such as binary outcomes (e.g., \cite{b2}). The concept of similarity within the random forest context is seen in \cite{b3} for regression and classification. In \cite{b4}, the author applied the case-specific random forests method of \cite{b3} to a dataset for a binary response from the Medical Information Mart for Intensive Care (MIMIC-II) database (\cite{b5, b6}). 

In survival analysis, one notion of similarity is seen in cure models. These models assume that while some cases will die from a disease or experimental stress, a sub-population will survive for a long time without experiencing the event. Although the term similarity is not specifically mentioned in this literature, the sub-population of long-term survivors can be considered as a group of similar cases. Early studies on such models include \cite{b7}, \cite{b8}, and \cite{b9}.  In \cite{b10}, the authors suggested a straightforward computational method to deal with grouped survival data based on the Cox proportional-hazards model. In both \cite{b11} and \cite{b12}, the respective authors used a mixture model representation for the two populations, which models the probability of being a long-term survivor with a logistic regression and the time to event for those that would experience the event with survival models, respectively. Many variations of mixture cure models can be seen in literature. In \cite{b13}, the authors provided an alternative to two-component mixture models in estimating cure rate by using bounded cumulative hazard function. These models focus on modeling rather than prediction. 

We take a rather different approach to model and predict survival data when there are one or more sub-populations in the dataset, that is, when the relationship between the time-to-event outcome and the explanatory variables are homogeneous within groups and more heterogeneous between groups. This is a more general case than the cure model as there can be more than two groups in the population, and the number of groups is unknown, in general. Note that the similarity is not just based on the grouping of the survival time, or the closeness of the explanatory variables, but depends on the relationship between the two. Tree-based methods such as random forests \cite{b14} are a natural way of incorporating both outcome and covariate information, and can be utilized to characterize similarity as cases in the same terminal node can be considered as similar to each other. Random forests methods have been extended to survival data as well, as in \cite{b15}, and our approach is essentially combining the case-specific random forests model in \cite{b3} with the random survival forests model \cite{b15}. An approach for handling dependent right censoring will be proposed as well.

In Section 2, we will discuss our proposed similarity-based random survival forest algorithm with independent right censoring, and methods to adjust for dependent censoring. Time-varying area under the receiver operating characteristic curve (time-varying AUC; note AUC is sometimes written as AUROC in the literature) is used as our primary criterion for evaluating prediction performance. In Sections 3 and 4, respectively, we present applications of the algorithm in a simulation study, as well to a real dataset from the MIMIC-III database \cite{b16}, an update to MIMIC-II (\cite{b5, b6}). In Section 5, we will summarize our methodology and findings from the simulation study and real data analysis.

\section{Similarity-based Random Survival Forest}

In this section, we will introduce the algorithm for our proposed similarity-based random survival forest (SB-RSF). The idea is to build a different random survival forest for prediction for each test case, giving greater weight to the training cases that are in closer proximity to the test case, and using less information from those that merely add more noise to prediction. We will discuss the methods under the assumption of independent censoring in Section 2.1 and then under the more flexible assumption of dependent censoring in Section 2.2. In Section 2.3, we will talk about using time-varying AUC for model comparison. 

\subsection{With Independent Censoring}

We will assume independent censoring for now. Methods to incorporate dependent censoring will be discussed in Section 2.2.

\begin{itemize}
\item 1. Construct a regular random survival forest model for a training dataset that has sample size $N_{train}$. 
	\begin{itemize}
		\item (a) Draw $B$ bootstrap samples from the training data. Uniform sampling is used.  
		\item (b) Grow a survival tree for each bootstrap sample under the constraint that it should have $d_0>0$ unique deaths. 
	\end{itemize}
\item 2.  For each point in the test dataset of size $N_{test}$, obtain a weight vector based on the random survival forest in the first step.
	\begin{itemize}
		\item (a) Pass a test data point down each tree in the random survival forest, and keep track of how many terminal nodes group a training data point with the test point. 
		\item (b) Assign a weight vector of length $N_{train}$ to each test data point based on how many terminal nodes group a training data with that test data point.
		\item (c) Iterate through each test data point, and obtain a weight matrix of size $N_{train} \times N_{test}$. Normalize each row of the weight matrix so that each row sums to 1.
	\end{itemize}
\item 3. Build a different similarity-based random survival forest for each test data point.
	\begin{itemize}
			\item (a) For a given test data point, build a random survival forest model with the weight vector as the sampling probability vector in the bootstrap. 
			\item (b) Pass down the test data point in each tree, and estimate the cumulative hazard function (CHF) of the terminal node to which the test data point belongs.
			\item (c) Average among all trees to get an ensemble CHF for that test data point.
			\item (d) Repeat (3.a)-(3.c) for each data point in the test dataset.
	\end{itemize}
\end{itemize}
 
\subsection{Adjusting for Dependent Censoring}
Dependent censoring for right-censored data is common in follow-up studies. For right censoring, the event is only known to have occurred after a certain time point. Denoting the censoring time by $C_i$, the observed time $X_i$ will be the minimum of the event time and the censoring time, i.e., $X_i=min(C_i, T_i)$. Denote the event indicator by $\delta_i$, which indicates the observed time corresponds to the true event time, then $\delta_i = I(T_i \leq C_i)$, which is 1 if the event occurs before censoring, and 0 otherwise. For non-informative censoring, the censoring process does not directly depend on the event process, although it can depend on some covariates. With informative censoring, the censoring process directly relates to the expected time to event. Inverse probability-of-censoring weights (IPCW) have been shown to account for the bias that occurs when ignoring informative censoring (\cite{b17, b18}).
In this setting, the algorithm is modified as follows:
\begin{itemize}
		\item 1.  Use the standard Kaplan-Meier estimator with censoring time as the event time to get the probability $P_{i}(C)$ of getting censored.
		\item 2.  Calculate the IPC weights for each training case as $IPCW_{i}=1/(1-P_{i}(C))$, i.e. the weights are equal to the inverse probability of not getting censored. 
		\item 3.  Calculate the similarity weights for a training case $i$ and test case $j$ as $SW_{i,j}$, as described in Section 2.1.
		\item 4. The sampling weights under dependent censoring for use in the similarity-based random survival forest for $i$ and $j$ will be proportional to $SamplingW_{i,j}=IPCW_{i}*SW_{i,j}$.
	\end{itemize}
The intuition behind the multiplication of the weights is that the SB-RSF algorithm now gives greater sampling weights to those data points that are more likely to be censored.

\subsection{Prediction Accuracy}
We will be using time-varying area-under-the receiver operating characteristic curve (time-varying $AUC$; sometimes written as time-varying $AUROC$) for model comparison. For binary outcomes, the prediction accuracy can be characterized by ROC, which plots the sensitivity against (1-specificity) for the range of possible decision-cutoff thresholds. And the area under ROC ($AUC$ or $AUROC$) represents a measure of prediction accuracy. 

For time-to-event outcomes, there are a few proposals to generalize the concept of sensitivity and specificity (e.g., \cite{b19}). One way is to look at sensitivity and specificity at each time of interest $t$. The survival probability up to $t$ of a test case $i$, i.e., $S_{i}(t)$, can be derived from its cumulative hazard $\hat{H}_{i}(t)$. Then, $AUC(t)$ can be estimated at each $t$; this is the time-varying $AUC$. In this paper, we will evaluate time-varying $AUC$ over a dense grid of time points.

\section{Simulations}
We use two simulated examples to further explain what similarity means in the model and demonstrate the prediction performance of the algorithm. 

\subsection{Example 1}
\begin{figure}
\vspace{0.1in}
\centering
\includegraphics[height=4in,width=4in, keepaspectratio]{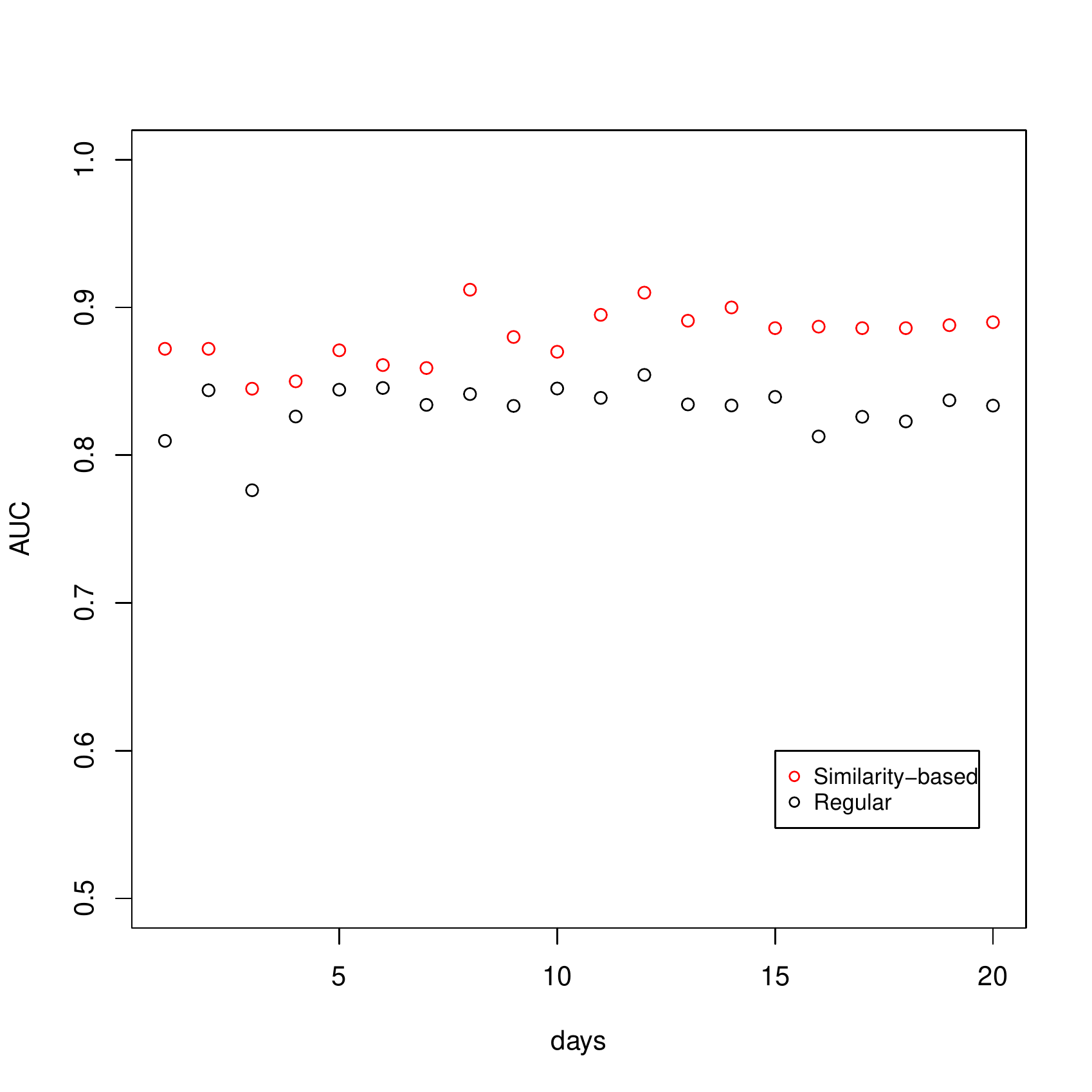} 
\caption{Time-varying $AUC$ for simulated data in Example 1}
\label{simex1}
\end{figure}

In a simple example, each case has a 3-dimensional covariate $\{X_1, X_2, X_3\}$ that links directly to the survival outcome. Two of the covariates are linked to similarity as well. In this case, $S$ is a survival outcome that follows a Weibull distribution with shape=2, and log(scale) mapped to linear predictor $Y$: 
%\begin{gather}
\begin{equation}
Y = 
\begin{cases}
0.2 X_1 - 0.1X_2+ 0.5X_3, & \hspace{-.25cm} \text{if }(X_1+7)*(X_3-10)>0 \\
0.3 X_1+0.1X_2 - 0.3X_3,  & \hspace{-.25cm} \text{otherwise} \ .
\end{cases}
\end{equation}
%\end{gather}
Here, $(X_1+7)*(X_3-10)<=0$ and $(X_1+7)*(X_3-10)>0$ describes a binary tree structure that clusters cases into two subspaces. Within each subspace, the relationship between the survival outcome and the covariates are the same, but different between the subspaces. 1000 cases are generated, where $X_1, X_2, X_3$ are independently and uniformly generated from (-15,15). Uniform right censoring (independent for now) is considered. Fig. \ref{simex1} summarizes the comparison between the prediction performance of case-specific random survival forest and the regular random survival forest. The red dots represent the time-varying $AUC$ for the case-specific random survival forest and the black dots are for the regular random survival forest. The $AUC$s are evaluated at each day from day 1 to day 20. At each day, the time-varying $AUC$ of the case-specific method exceeds the regular random survival forest, more often than not by a sizable margin.
 
\subsection{Example 2}
In the second model, each case has a 5-dimensional covariate $\{X_1, X_2, X_3, X_4, X_5\}$, where three of the covariates explain similarity. Again, we will use a binary tree structure to define subspaces. In this case, we will prune the tree until there are four terminal nodes, i.e., four subspaces. Again, within each subspace, the relationship between $Y$ and the covariates are the same, but different between subspaces. 

\begin{figure}
\vspace{0.1in}
\centering
\includegraphics[height=4in,width=4in, keepaspectratio]{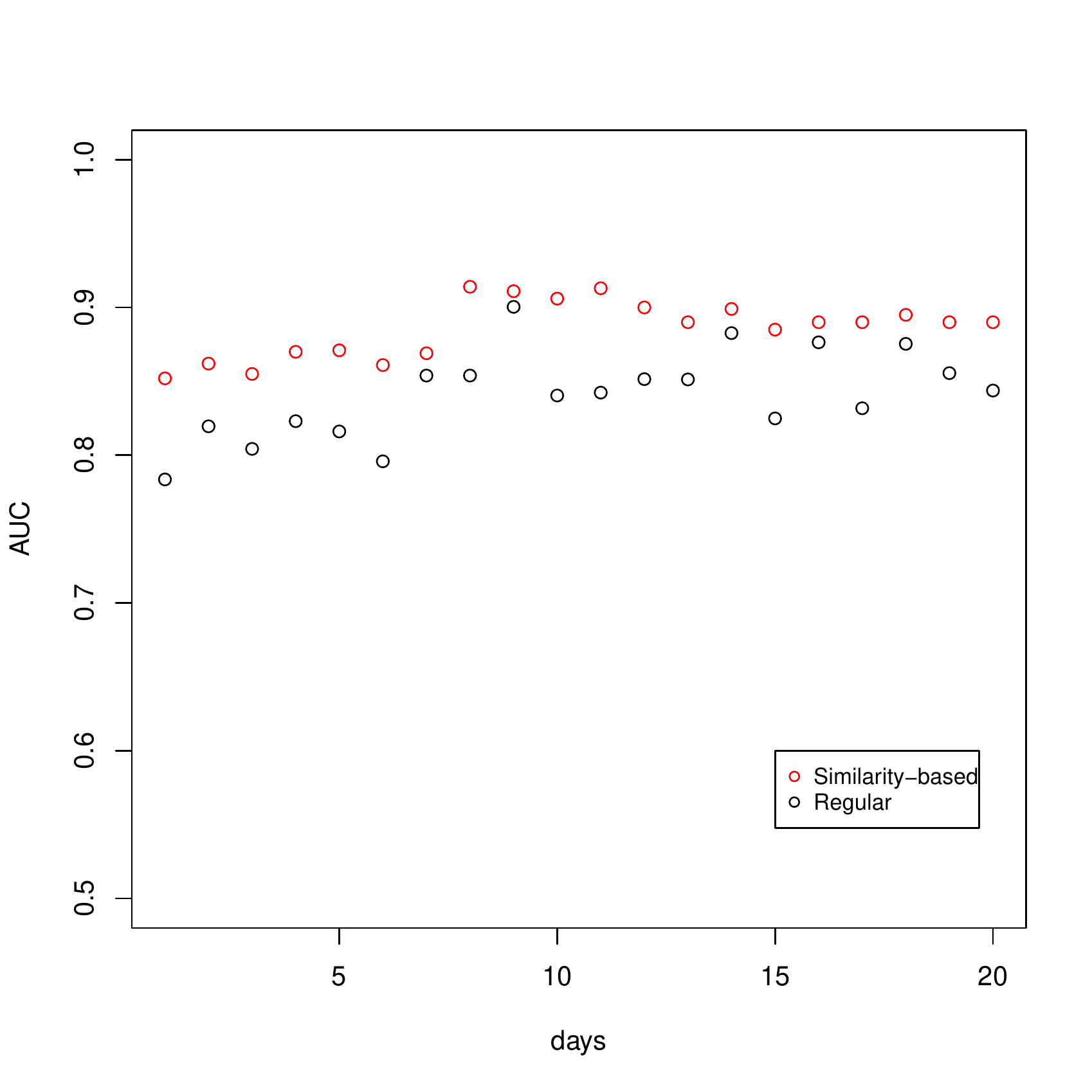} 
\caption{Time-varying $AUC$ for simulated data in Example 2}
\label{simex2}
\end{figure}

The result in Fig. \ref{simex2} is similar to the first simulation result in Fig. \ref{simex1}.  Giving more weights in the sampling to similar cases, based on our SB-RSF method, yields better predictive accuracy in the random survival forest framework.

\section{Application to an ICU dataset}

\subsection{MIMIC-III}
MIMIC-III (Medical Information Mart for Intensive Care III) is a freely accessible critical care database for 53,423 distinct hospital admissions for adult patients (aged 16 and above). Data includes vital signs, medications, diagnostic code, survival data and high resolution data including lab results and bedside monitoring data \cite{b16}. 

This large dataset provides rich information for modeling and prediction, but the diversity of the patients also poses challenges to accurate prediction of outcome of interest. To illustrate, the goal is to predict ICU patient survival with their age, gender, ICU type, admission type, and severity of disease classification score, SAPS II \cite{b20}, as predictors. ICU type includes CCU (Coronary Care Unit), CSRU (Cardiovascular Intensive Care Unit), MICU (Medical Intensive Care Unit), SICU (Surgical Intensive Care Unit) and TSICU (Trauma Surgical Intensive Care Unit). Admission type includes Elective, Emergency, and Urgent. Only the first hospital admission of adult patients (older than 15 years of age) is included in our study. Excluding cases with missing data in one or more of the variables or outcome, the sample size is 38,604. In this dataset, 80\% of the cases are right-censored at 90 days after hospital discharge, for the purpose of de-identification. 

\subsection{Result}
\begin{figure*}
\vspace{0.1in}
\centering
\includegraphics[height=6in,width=7in, keepaspectratio]{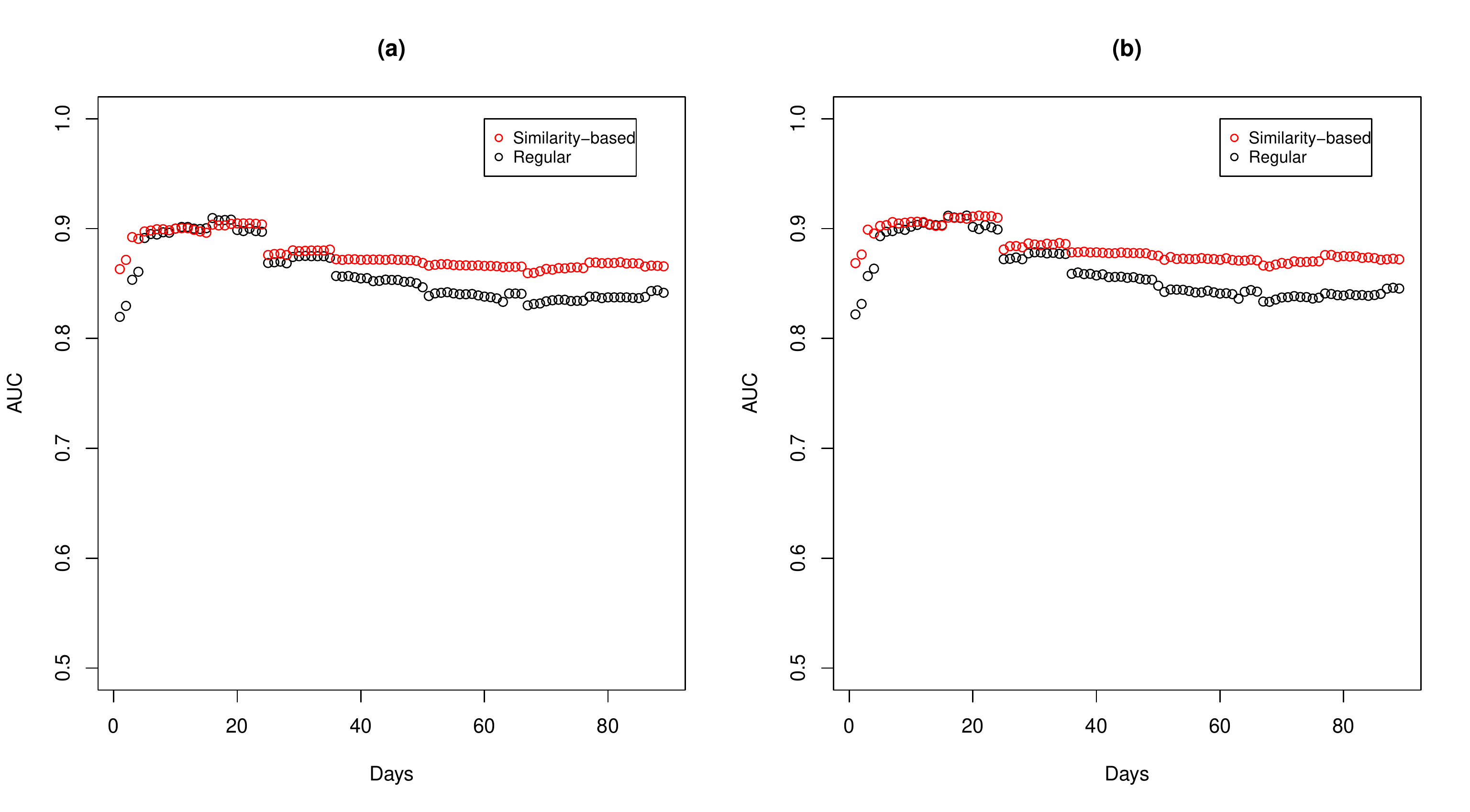} 
\caption{Time-varying $AUC$ for application to MIMIC III dataset. (a) Ignoring the dependency in censoring (b) Adjusted for dependent censoring}
\label{AUCreal}
\end{figure*}

Fig. \ref{AUCreal}(a) compares the time-varying $AUC$ for the algorithm in Section 2.1 with the random survival forests method. The time-varying $AUC$ from our proposed SB-RSF method outperforms that of the regular random survival forest at the beginning of the prediction and after day 20, and the gap between the two lines increases as we predict further into the future. 

Fig. \ref{AUCreal}(b) shows the result when considering possible dependency in the censoring. The result is similar to that in Fig. \ref{AUCreal}(a). It is possible that for this dataset there is not much dependency in the censoring, and thus the calculation of the IPC weights did not have a big impact on the result.

\section{Discussion}
In this paper we proposed to improve the random survival forests by incorporating the similarity structure between a test data point and training data point. Instead of building a global random survival forests for each test case, we construct similarity-based random survival forests for each one of them, by giving more weights to the training cases that are in closer proximity to the test case. Proximity is measured using a regular random survival forests model. We also developed an algorithm to account for dependent censoring which is common in survival data. 

Both simulations and a real data example show promising results that, in general, indicate that the similarity-based prediction improves predictive performance of random survival forests in terms of time-varying $AUC$. This result is also consistent with other findings using similarity structure for binary response data (e.g., \cite{b2}). 

Our proposed SB-RSF method requires building a random survival forest for every test data point and specification of a few tuning parameters. Specifically, the tuning parameters are the depth of the tree (represented by the number of unique deaths in the terminal nodes), the number of candidate predictors to consider for splitting at each node, and the number of trees in the forests. This leads to a computationally intensive algorithm, especially when the size of the test data size is large. Future work to investigate ways in alleviating some of this computational burden would be helpful.
%Future work is necessary to investigate the robustness of their choices. 
One way of reducing computation time is to use a hard threshold for sampling, that is, giving 0 weight to cases that are too far away from the test case. The tuning parameters for the simulations are selected based on the entire training dataset. However, if they are determined from a smaller subset of the training data, the computational time might be greatly reduced.

For future work, methods other than random forests may be utilized for similarity-based prediction for survival outcomes. One possible extension is the joint modeling of longitudinal covariates and a time-to-event outcome (e.g., \cite{b21}). One might be able to identify similar cases based on longitudinal covariates as well as time-fixed covariates.  In addition, an approach within this framework that handles missing values in the dataset should be pursued as well.  

In spite of some areas that require future study, we have shown the proposed SB-RSF approach to hold promise for the prediction of survival outcomes. Our investigation shows that our similarity-based algorithm can improve the predictive accuracy of a popular and useful prediction tool, i.e., random survival forest (\cite{b15, b22}), for time-to-event data.

%\vspace{12pt}
%\color{red}
%IEEE conference templates contain guidance text for composing and formatting conference papers. Please ensure that %all template text is removed from your conference paper prior to submission to the conference. Failure to remove the %template text from your paper may result in your paper not being published.

\end{document}